\documentclass[12pt,preprint]{aastex}

\usepackage{lscape}

\newcommand{\etal}{et al.}      
\newcommand{\inv}{$^{-1}$}
\newcommand{\kms}{km~s\inv}

\newcommand{\oii}{[\ion{O}{2}] $\lambda 3727$~}
\newcommand{\oiii}{[\ion{O}{3}]}
\newcommand{\hbeta}{H$\beta$}

\newcommand{\Rvirial}{R_{200}}

\slugcomment{Draft 1 2011-May-2}

\slugcomment{}

\shorttitle{LCBGs in Distant Clusters}
\shortauthors{Crawford et al.}

\begin{document}


\title{Spectroscopy of Luminous Compact Blue Galaxies\\in Distant 
Clusters I. Spectroscopic Data\altaffilmark{1}}

\author{Steven M. Crawford}
\affil{SAAO, P. O. Box 9, Observatory 7935, Cape Town, South Africa}
\email{crawford@saao.ac.za}

\author{Gregory D. Wirth}
\affil{W. M. Keck Observatory, 65-1120 Mamalahoa Hwy, Kamuela HI 96743}
\email{wirth@keck.hawaii.edu}

\author{Matthew A. Bershady}
\affil{Department of Astronomy, University of Wisconsin, 475 North Charter Street, Madison, WI 53706}
\email{mab@astro.wisc.edu}

\and

\author{Kimo Hon}
\affil{W. M. Keck Observatory, 65-1120 Mamalahoa Hwy, Kamuela HI 96743}
\email{hon.kimo@gmail.com}

\altaffiltext{1}{Based in part on data obtained at the W. M. Keck
  Observatory, which is operated as a scientific partnership among the
  California Institute of Technology, the University of California,
  and NASA, and was made possible by the generous financial support of
  the W. M. Keck Foundation.}




\begin{abstract}

  We used the DEIMOS spectrograph on the Keck II Telescope to obtain
  spectra of galaxies in the fields of five distant, rich galaxy
  clusters over the redshift range $0.5 < z < 0.9$ in a search for
  luminous, compact, blue galaxies (LCBGs).  Unlike traditional
  studies of galaxy clusters, we preferentially targeted blue cluster
  members identified via multi-band photometric pre-selection based on
  imaging data from the WIYN telescope.  Of the 1288 sources that we
  targeted, we determined secure spectroscopic redshifts for 848
  sources, yielding a total success rate of $66\%$.  Our redshift
  measurements are in good agreement with those previously reported in
  the literature, except for 11 targets which we believe were
  previously in error.  Within our sample, we confirm the presence of
  53 LCBGs in the five galaxy clusters. The clusters all stand out as
  distinct peaks in the redshift distribution of LCBGs with the
  average number density of LCBGs ranging from
  $1.65\pm0.25~\mbox{Mpc}^{-3}$ at $z=0.55$ to
  $3.13\pm0.65~\mbox{Mpc}^{-3}$ at $z=0.8$.  The number density of
  LCBGs in clustes exceeds the field desnity by a factor of
  $749\pm116$ at $z=0.55$; at $z=0.8$, the corresponding ratio is
  $E=416\pm95$.  At $z=0.55$, this enhancement is well above that seen
  for blue galaxies or the overall cluster population, indicating that
  LCBGs are preferentially triggered in high-density environments at
  intermediate redshifts.

\end{abstract}





\section{Introduction}\label{sxn-intro}

The first provocative evidence of galaxy evolution in the Universe was
the increasing fraction of blue galaxies in clusters reported in the
now-classic papers by Butcher \& Oemler (1978, 1984).  In these early
papers, based purely on photometry, the large scatter in the blue
fraction from cluster to cluster -- along with some counter examples of
very red clusters at what was then considered ``high'' redshift (e.g.
Cl 0016+16 by Koo 1981) -- made it unclear just how rapidly and
uniformly cluster populations were evolving.  Despite the passage of
three decades since the first publication of these papers, we still
lack a definitive picture of how star-forming populations in galaxy
clusters evolve.  The situation among cluster galaxies stands in stark
contrast to the substantial evolution observed in the field galaxy
population, in which abundant redshift surveys have now revealed a
rapid rise in the star formation rate to $z=1$ (Cooper \etal\ 2008). A
major impediment to improving the understanding of cluster evolution
has been a lack of studies probing the star-forming populations in
clusters, especially at intermediate redshifts ($0.3 < z < 1.0$).

The first confirmation of blue, cluster galaxies was by Dressler \&
Gunn (1982); they used spectroscopic observations to confirm the
cluster membership of the objects and explore their properties.
Further spectroscopic observations of clusters indicated a group of
transitional objects (e.g., ``E+A'' galaxies, Dressler \& Gunn 1983)
that could provide a link between actively star-forming objects (in
what is today referred to as the ``blue cloud'') and passive galaxies
lying on the ``red sequence'' (Couch \& Sharples 1987, Wirth \etal\
1994, Barger et al. 1996, Tran \etal\ 2003).

The stark difference between the cluster and the field populations --
as described by the morphology-density relationship (Dressler \etal\
1980) or the star formation-density relationship (Gomez \etal\ 2003)
-- prompted investigators to invoke numerous mechanisms that would
lead to an environmental dependence in galaxy formation and evolution.
In a hierarchical formation model, galaxies falling into the cluster
environment are transformed into a quiescent population through a
variety of mechanisms that extinguish star formation via gas
starvation, stripping, and/or pre-processing (see the review by
Boselli and Gavazzi 2006).  Evidence for these different processes has
been observed at both low and high redshift, and these transformations
apparently start well outside the cluster virial radius (Porter \&
Raychaudhury 2005, Poggianti \etal\ 2009).

One of the first attempts to produce a comprehensive inventory of
star-forming population of an intermediate redshift cluster was a
narrow-band imaging survey by Martin, Lotz, and Ferguson (2000) of
Abell 851 at $z=0.45$ targeting \oii emission.  They reported an
overabundance of star-forming galaxies in the clusters as compared to
the field at similar redshift, but this result was not confirmed by
their subsequent observations of the lower mass cluster MS 1512.4+3647
at $z=0.372$ (Lotz, Martin, and Ferguson 2003).  However, Abell 851 is
a far more massive cluster with significant evidence of substructure,
so differences between the star-forming populations may be related to
the different properties and evolutionary states of the clusters.
Finn \etal\ (2004) expanded on these measurements with H$\alpha$
narrow-band observations of intermediate-redshift clusters, finding an
increase in the total star formation rate with increasing redshift
among the cluster population (Finn \etal\ 2008) that matches the
increase which is found in the global star formation rate (Madau
\etal\ 1997, Cooper \etal\ 2008).

The presence of obscured star-forming galaxies further complicates the
picture of star formation in clusters.  Radio continuum observations
provided the first evidence for the presence of these sources (Miller
\& Owen 2002), and a series of studies at far-infrared wavelengths has
found a large number of heavily-obscured star-forming galaxies
(Saintonge, Tran, and Holden 2008; Gallazzi \etal\ 2009; Haines \etal\
2009).  In follow-up spectroscopy to their narrow-band observations of
A851, Sato \& Martin (2006a, 2006b) identified a population of
heavily-reddened, star-forming galaxies and bursting dwarf
populations.  It remains a challenge to explain the properties of
these objects and how they pertain to the evolution of cluster
galaxies (Smith \etal\ 2010), and in particular, the connection, if
any, to the star-forming population that is {\it not} heavily
obscured.

To illuminate the interplay between environment and evolution, a large
number of recent spectroscopic surveys has targeted clusters at
intermediate redshift (Postman \etal\  2001, Tran \etal\  2003, Halliday
\etal\  2004, Sato \& Martin 2006a, Moran \etal\  2007, Tanaka et
al. 2007).  Only with this added kinematic information can we
determine whether we are seeing galaxies that are falling into the
cluster for the first time, a backsplash population of objects
(Pimbblet 2011), or objects forming in-situ in the cluster such as
tidal dwarfs (Duc \& Bournaud 2008). By extending the kinematic
coverage to a comprehensive sample of the cluster star-forming
galaxies, we can then hope to establish a clear connection between
these star-forming galaxies at intermediate redshift and the
populations seen in clusters today.

In this paper, we specifically focus on Luminous Compact Blue Galaxies
(LCBGs), an extreme star-forming class of galaxies initially
identified in the field at intermediate redshifts (Koo \etal\ 1994).
Their sharp drop in number density with decreasing redshift mimics the
decline in the global star formation rate (Guzman \etal\ 1997, Werk
\etal\ 2004), and the population appears to be a heterogeneous mix of
bursting dwarfs and star-forming bulges (Guzman \etal\ 1996, Garland
\etal\ 2004, Noeske \etal\ 2006, Rawat \etal\ 2007, Tollerud \etal\
2010).  Due to this observed mix, LCBGs are proposed either to evolve
into spheroidal systems\footnote{For our purposes, we refer to
  spheroidal systems as either dwarf spheriodals or dwarf ellipticals
  or other similar low mass systems.} (Koo \etal\ 1994, Guzman \etal\
1996) or to be an intermediate phase in the evolution of
bulge-dominated spiral galaxies (Phillips et al. 1997, Hammer \etal\
2001).

Follow-up observations of a small sample of blue galaxies in Cl
0024+1654 at $z=0.39$ by Koo \etal\ (1997) provided the first
confirmation of LCBGs in galaxy clusters.  Crawford \etal\ (2006)
found an enhancement in these types of galaxies among intermediate
redshift clusters based purely on photometric information, thus
suggesting an increase in both their number density and fraction of
galaxies with galaxy density.  Further spectroscopic confirmation of
cluster LCBGs was reported by Moran et al. (2007) in MS 0451-03 at
$z=0.54$.

In this paper, we introduce our survey and present optical
spectroscopic measurements obtained from two observing runs with the
DEIMOS spectrograph on the Keck II Telescope.  In \S \ref{sxn-wltv},
we describe the WIYN Long Term Variability survey from which our
sample was selected.  In \S \ref{sxn-spec}, the spectroscopic
observations from sample selection to data reduction are presented.
In \S \ref{sxn-objects}, we provide the catalog of targeted objects.
We discuss the quality of our redshift measurements in \S
\ref{sxn-analysis}.  Finally, we briefly examine the evolution of
cluster LCBGs with environment and redshift in \S \ref{sxn-lcbgs}.

Throughout this work, we adopt $H_{0}=70$ km s$^{-1}$ Mpc$^{-1}$,
$\Omega_{M} = 0.3$, and $\Omega_{\Lambda} = 0.7$; all magnitudes
are in the Vega system.

\section{The WLTV Survey}\label{sxn-wltv}

The WIYN Long Term Variability (WLTV) survey is a photometric census
of ten massive galaxy clusters over the redshift range $0.3 < z < 0.9$
undertaken with the WIYN 3.5~m telescope.  The observational aim of
the survey was to acquire deep, multi-epoch photometry from the
near-UV to the near-IR in very rich clusters at intermediate
redshifts.  The observations were completed over a 6-year period and
sample the time domain on scales of one month up to the survey
duration.  Our extragalactic scientific goals include detailed star
formation and stellar population studies of individual cluster
galaxies; cluster populations as well as galaxies in the foreground,
background, and cluster outskirts; and a search for transients
(supernovae) and AGN variability in galaxies within the field of rich
clusters.  The photometric band-passes and depth chosen to achieve
these goals are described in \S \ref{sxn-imaging}.

\subsection{Cluster Sample}\label{sxn-sample}

As detailed in Crawford \etal\ (2009), we established the following
three key criteria to select clusters for the survey:

\begin{enumerate}

\item general recognition in the literature that the cluster
  represents a significant, high-redshift overdensity in the galaxy
  distribution;

\item availability of Hubble Space Telescope (HST) imaging data of 
the field to permit accurate measurements of galaxy size and
morphology;

\item existence of significant followup spectroscopy establishing the 
overdensity as a bona-fide cluster rather than a chance superposition.

\end{enumerate}

Since the start of the WLTV observing campaign in 1999, many of these
clusters have been observed by others across a wide range of
wavelengths.  From this sample, we have selected the five
highest-redshift, most massive clusters for further investigation.
Details of the five selected clusters\footnote{A sixth high-density
  region was originally targeted, but follow-up imaging observations
  indicated that it was not a bona fide cluster. This region is
  adjacent to the Cl 1324+3011 observations.} are provided in Table
\ref{tab_cluster_params}.  In this table, we provide the cluster name,
the survey identifier for each cluster, redshift, cluster velocity
dispersion, $M_{200}$ and $R_{200}$ radius\footnote{$M_{200}$ and
  $R_{200}$ are based on the definition from Finn \etal\ (2005).} for
each cluster.  The cluster velocity dispersion is calculated based on
all available spectroscopic data following a method similar to Fadda
\etal\ (1996), and the full details of the calculations will be given
in future work.  A description of each of the major clusters is
provided below:

\begin{itemize}

\item {\bf MS 0451-03} is a rich, well-studied cluster at $z=0.53$
  initially discovered by Stocke \etal\ (1991) in the Einstein Medium
  Sensitivity Survey (EMSS) and spectroscopically confirmed by Gioia
  \& Luppino (1994).  MS 0451-03 is X-ray luminous (Donahue \etal\
  2003), with over 300 spectroscopically-confirmed members (Ellingson
  \etal\ 1998, Moran \etal\ 2007).  Cluster mass estimates are
  available from the velocity dispersion of cluster galaxies (Carlberg
  \etal\ 1996), X-ray luminosity (Donahue \etal\ 2003), and
  weak-lensing analysis (Clowe et al. 2000).

\item {\bf Cl 0016+16} is one of the first clusters found to defy the
  reported increase in the fraction of blue cluster galaxies at
  intermediate redshift (Koo 1981).  Specifically, the core of this
  elongated cluster has few blue galaxies.  Cl 0016+16 is a rich
  cluster at $z=0.55$ with over 200 spectroscopically confirmed
  members (Wirth \etal\ 1994, Ellingson et al. 1998, Dressler \etal\
  1999, Tanaka \etal\ 2007).  It is the major component of a
  supercluster complex (Connolly \etal\ 1996, Tanaka et al. 2005,
  2007).  X-ray observations of the cluster reveal a luminous system
  with multiple substructures (Worrall \& Birkinshaw 2003), but yield
  mass estimates comparable to other measurements based on galaxy
  velocity dispersion and weak lensing (Smail \etal\ 1997, Carlberg
  \etal\ 1997, and Clowe \etal\ 2000).

\item {\bf Cl J1324+3011} was originally discovered by Gunn, Hoessel,
  \& Oke (1986) and spectroscopically confirmed as a cluster at
  $z=0.75$ by Oke, Postman \& Lubin (1998).  XMM-Newton observations
  of the cluster indicate it is under-luminous for its velocity
  dispersion as compared to local galaxy clusters (Lubin, Mulchaey, \&
  Postman 2004).


\item {\bf MS 1054-03} is a massive cluster at $z=0.83$ that has been
extensively studied both through HST imaging and spectroscopy (van
Dokkum \etal\ 1999, Tran \etal\ 1999, Goto \etal\ 2005, Tran et
al. 2005).  Initially discovered as one of the highest-redshift
sources in the EMSS (Stocke \etal\ 1991), MS 1054-03 has been shown to
be a massive cluster at high redshift on the basis of spectroscopic
velocity dispersion measurements (Tran \etal\ 1999), X-ray luminosity
(Donahue \etal\ 1998, Neumann \& Arnaud 2000, Jeltema et al 2001,
Gioia \etal\ 2004), and weak-lensing mass estimates (Luppino \& Kaiser
1997, Clowe \etal\ 2000, Jee \etal\ 2005).

\item {\bf Cl J1604+4304} is the highest-redshift cluster in
our sample at $z=0.90$ (Oke, Postman, Lubin 1998).  It forms part of a
supercluster complex (Gal \& Lubin 2004) and exhibits an overdensity
of AGN (Kocevski \etal\ 2009). The X-ray luminosity of the cluster is
lower than predicted from the measured velocity dispersion (Lubin et
al. 2004).  The uncertainty in the mass from a weak-lensing estimate
does not allow strong constraints on the cluster mass, but does
confirm the presence of a massive structure at high redshift
(Margoniner et al. 2005).

\end{itemize}

All of the clusters in our sample are massive, and due to their
predicted growth, they are likely to all have similar mass to each
other if observed today.  Following the models of Wechsler \etal\
(2002) for the growth of dark matter structures, we would predict
these structures to have a velocity dispersion $\sigma \sim 1500$ km
s$^{-1}$ and masses of $6 \times 10^{15} M_{\odot}$ at the present
epoch.

\subsection{Imaging Survey}\label{sxn-imaging}

The core of the time-domain WLTV imaging survey consisted of $UBRI$
imaging with the Mini-Mosaic camera ($10\arcmin$ field of view with
$0\farcs14~\mbox{px}^{-1}$) on the WIYN 3.5~m telescope over six years
from October 1999 until June 2005. For the purposes of deriving
photometric redshifts and rest-frame $B$-band properties of the
highest-redshift cluster galaxies, the data were supplemented with
deep $z$-band imaging that we obtained at WIYN with the same
instrument.  The typical limiting magnitude of each field is
$R\sim25.5$, with similar depth in the other passbands.  Full details
of the observations, data reduction, and analysis appear in Crawford
\etal\ (2009).

For the highest-redshift clusters among the sample, we designed a set
of custom narrow-band filters to observe the \oii\ spectral feature in
star-forming galaxies at the redshift of each cluster.  The width of
each narrow-band filter was set by the velocity dispersion of the
cluster and is typically $\sim100$ \AA.  Observations were obtained
through the narrow-band filters for a minimum of 3.5~h using the same
instrumentation and telescope as for the broad-band imaging program.
We also observed each cluster with an off-band narrow-band filter that
is close to the on-band filter, but sufficiently different in central
wavelength to avoid contamination from cluster sources.  Measurements
of the strength of the \oii\ feature were derived from fits to the
full spectral energy distribution and are used in our selection of
spectroscopic targets.  Full details of the narrow-band filters and
data reduction are presented in Crawford (2006).

\section{Spectroscopic Observations}\label{sxn-spec}

\subsection{Sample Selection \& Masks}\label{sxn-masks}
Since the aim of our present investigation is to identify star-forming
cluster galaxies, we deviated from the customary strategy for
observing high-redshift clusters by preferentially selecting blue
(rather than red) cluster objects for spectroscopy.  These targets
were selected on the basis of photometric measurements derived from
the WLTV narrow-band survey data.  Potential emission-line galaxies
were identified via a flux excess in the on-band filter combined with
the estimated photometric redshift.  Due to improvements in the
technique of measuring the flux excess, our selection criteria
differed between the two Keck observing runs as described below.

For the November 2005 run which included Cl 0016+16 and MS0451-03, we
assigned top priority to objects classified as LCBGs.  As further
discussed in \S \ref{sxn-class}, we adopted the definition from
Crawford \etal\ (2006) with LCBGs defined as galaxies with
$(B-V)_o<0.5$, $\mu_B < 21$ \mbox{mag arcsec}$^{-2}$, and $M_B <
-18.5$.  Next highest priority was given to other cluster star-forming
galaxies; i.e., objects showing blue colors and an excess in the
narrow-band filter sampling \oii\ at the cluster redshift as compared
to the continuum filter.  Specifically, these blue objects were
defined as having $B-I < 2.5$ mag and $C-E > 0.2$ mag where $E$ is the
measured flux within the \oii\ filter for each cluster and $C$ is the
flux within the corresponding blueward continuum filter.  The apparent
color of $B-I=2.5$ would correspond to having a rest-frame color of
$(B-V)_o\sim0.5$ at $z=0.55$.  For each class of objects, higher
priority for selection was granted to sources with a spectroscopic or
photometric redshift within $|\Delta z| \leq 0.1$ of the nominal
cluster redshift.  Finally, brighter galaxies were given higher
selection priority to maximize the resulting number of usable spectra.
We applied an apparent magnitude cut at $R<24.0$ and rejected bright
stars (defined as having $R<22.5$ and a half-light radius of
$r_{0.5}<0\farcs5$).  Because each DEIMOS mask covers an area much
larger than the WIYN field of view, we used $R$-band pre-imaging
obtained with DEIMOS to select additional targets for spectroscopy in
areas outside the WLTV survey field.  These objects were selected
purely based on their $R$-band magnitude with preference given to
brighter objects.

For the April 2007 run which included MS 1054-03, Cl J1324+3011, and
Cl 1604+4304; we modified the selection criteria to include
information from the improved determination of the narrow-band flux.
We preferentially selected cluster emission-line objects, identified
as having a high probability of cluster membership based on on-band
flux, color, and photometric redshift.  The on-band flux was
determined by fitting the full observed Spectral Energy Distribution
(SED) with model SEDs and then subtracting off the continuum value at
the on-band filter. We assigned the next highest priority to potential
high-redshift QSOs and Ly~$\alpha$ galaxies, respectively, both of
which were selected based on their colors using the Lyman break
technique (Guhathakurta, Tyson, and Majewski 1990; Cowie \& Hu 1998).
Next, preference was given to blue cluster objects, additional cluster
objects, and non-cluster blue objects.  Due to the higher redshift of
the clusters, we applied a fainter limiting magnitude of $R<25.0$ and
rejected bright stars from the target list.  All objects were selected
from the WIYN field of view as there was no pre-imaging available in
these fields.

To design slitmasks for DEIMOS we employed the
\textsc{DSIMULATOR}\footnote{\url{http://www.ucolick.org/\~{}phillips/deimos\_ref/masks.html}}
software provided by A. C. Phillips of UCO/Lick Observatory.  We
adjusted the placement of slits to maximize the number of potential
blue cluster objects on each mask.  For the Cl 0016+16 and MS0451-03
fields, astrometry was based on DEIMOS $R$-band pre-imaging over the
field of view.  For the other fields, astrometry was based on our WIYN
images and the sky position angle of the DEIMOS slitmasks was selected
to maximize the number of targets receiving slits on the mask.  The
position of the masks relative to the clusters can be seen on Figures
\ref{fig-radec-dss-w05}-\ref{fig-radec-dss-w10}.

\subsection{Observations}\label{sxn-obs}

We completed spectroscopic observations of the clusters fields using
DEIMOS on the Keck~II Telescope during 2005 November and 2007 April as
detailed in Table~\ref{tab-obslog}.  The observations comprised 15
slitmask fields with an average of 84 slits per mask.  We employed
different gratings, central wavelengths, and order-blocking filters in
order to maximize the likelihood of observing key diagnostic features
(chiefly \oii, \hbeta\ , and \oiii~$\lambda \lambda 4959,\ 5007$) at
the cluster redshift.  Each slitmask was observed for a total
on-source integration time of at least 3600~s, broken up into
$3\times1200$~s integrations to allow for the rejection of cosmic
rays. Two masks received an additional 1200~s of exposure in twilight.
No dithering took place between exposures because the masks employed
tilted slits and because the minor fringing pattern present in DEIMOS
images is sufficiently corrected by the use of flat field images.

For each mask we obtained a single arc spectrum including Na, Ar, Kr,
and Xe lamps to define the wavelength scale, and we acquired three
flatfield images using the internal halogen lamp to correct for minor
fringing and pixel-to-pixel sensitivity variations.  The closed-loop
flexure-compensation system of DEIMOS helps ensure that these
calibrations are spatially coincident with the on-sky spectra to
within $\pm0.25$~pixels even though the calibrations for the 2005 data
were acquired a month after the corresponding on-sky observations.
The seeing measured from stars in our slitmask alignment images was
typically in the range of $0\farcs8$--$1\farcs2$ (FWHM). Transparency
was generally good, although some minor cirrus affected the 2005
observations.

\subsection{Spectroscopic Reductions}\label{sxn-redux}

We reduced the spectra using the fully-automated DEIMOS data reduction
pipeline developed for the DEEP2 redshift survey (Davis et al. 2003,
Davis et al. 2007) and generously shared with us by the team (Newman,
private communication).  For each mask, the pipeline used the single
arc-lamp spectrum to define the wavelength scale for each mask and
used the flatfield images to derive corrections for CCD fringing and
pixel-to-pixel sensitivity variation.  The software combined the
multiple on-sky exposures into a single master image cleaned of cosmic
rays and removed the sky background from each slit by modeling the
night-sky emission with a fourth-order B-spline function (de Boor
1978) and subtracting the fit from the data to yield a 2-D
sky-subtracted spectrum.  The pipeline then produced a 1-D spectrum by
summing the flux within the illuminated pixels.  In the majority of
cases the pipeline worked well, but in a significant number of slits
the object spectrum did not appear in the position predicted by the
pipeline.  In such cases, the pipeline identified the desired spectrum
as a serendipitous target and extracted that spectrum as well.  We
corrected these misidentifications manually, as described below.

\subsection{Redshift Determination}\label{sxn-redshifts}

The process of determining redshifts and quality codes for each target
involved three phases.  First, we used an automated cross-correlation
technique to derive an estimated redshift for each target.  This
involved converting the 1-D spectra output by the DEEP2 pipeline, in
which the wavelength scale is irregular, to a linear wavelength scale
via linear interpolation.  In IRAF\footnote{IRAF is distributed by the
  National Optical Astronomy Observatory, which is operated by the
  Association of Universities for Research in Astronomy (AURA) under
  cooperative agreement with the National Science Foundation.}, we
employed the XCSAO task in the RVSAO radial velocity package (Kurtz \&
Mink 1998) to estimate the redshifts.  We selected 10
cross-correlation template spectra, supplied as part of the standard
RVSAO IRAF package, representing a variety of emission- and
absorption-line galaxy systems.  We found that if the estimated
starting redshift was off by more than $|\Delta z| > 0.1$ from the
actual redshift, XCSAO did not perform well; hence, we ran XCSAO
repeatedly with starting redshifts varying from $0.0 < z < 1.5$ at
intervals of $\delta z = 0.1$.  For each template, we selected the
redshift with the highest correlation coefficient as the best guess
for that template.  This process resulted in a set of 10 estimated
redshifts for each target, one per template.

The second phase involved having two or more reviewers manually
inspect each spectrum to determine the redshift and quality code.  Our
customized software package allowed the reviewer to select one of the
redshifts derived from the cross-correlation analysis, to select $z=0$
(star), to estimate a redshift manually by fitting to a spectral
feature, or to specify that no redshift could be determined.  We used
one of the cross-correlation redshifts whenever possible, but in cases
for which none of these automated redshift estimates was correct a
manual redshift based on a line fit was used instead.  The software
also allowed the reviewer to record the presence of key spectral
features and to note the presence of any one of a number of problems
which could affect the data.  Our redshift quality codes (hereafter,
$Q$; see Table~\ref{tab-zqual}) are the same as those employed in the
TKRS survey (Wirth et al. 2004).

The third phase involved reconciling any discrepant results from the
independent reviewers.  At this stage, one of us reviewed each
spectrum with discrepant redshifts, quality codes, or other
characteristics and made the final determination.  As a final step, we
manually inspected any spectrum which we suspected of being
misidentified as a serendipitous target, and modified the catalog to
correct the problem.

\section{Catalog and Classification of Cluster Objects}\label{sxn-objects}

\subsection{Object Catalog and On-Sky Distribution}

In Table \ref{tab-data}, we present the results from our spectroscopic
measurements (a full version appears online).  Information for all
sources targeted in our survey includes their measured redshift and
photometric classification. Redshifts are provided for all sources
with secure measurements.  The columns in Table \ref{tab-data} are:
(1) Identification in WLTV survey, (2) Right Ascension, (3)
Declination, (4) total $R$ magnitude, (5) mask name, (6) slit number,
(7) measured spectroscopic redshift, (8) redshift quality code, (9)
literature redshift, (10) reference, and (11) photometric
classification. Right Ascension and Declination are based on either
the DEIMOS pre-imaging or the WIYN imaging.  In both cases,
astrometric solutions for the images were determined from comparisons
to the USNO A2 catalog (Monet et al. 1998) with $0\farcs2$ rms.  Total
magnitudes are corrected for the shape of the source and are described
in Crawford et al. (2009).  Previously-measured spectroscopic
redshifts are listed and the reference for each redshift is provided.
The list of references is provided with the table.  Photometric
classifications are described in the next section.

In Figures \ref{fig-radec-dss-w05}-\ref{fig-radec-dss-w10}, we
present the projected sky distribution for our targets in each of our
fields.  Objects with secure spectroscopic measurements are indicated
by blue diamonds.  In each figure, we display an outline of the
respective fields of view for the WIYN imaging and the DEIMOS
spectroscopy along with the derived $R_{200}$ radius for the cluster.
The distribution in color-magnitude space for our sources with
successful spectroscopy can be seen in the left-hand panel of Figure
\ref{fig-master-class}.

\subsection{Galaxy Classification}\label{sxn-class}

For all sources with WIYN photometry, we provide a photometric
classification, which will be used in subsequent papers in this series
to differentiate between various cluster populations.  We divide
cluster sources into the following classes: Red Sequence (RS), Blue
Cloud (BC) and Luminous Compact Blue Galaxies (LCBGs).  While RS and
BC are exclusive classifications, LCBGs are a subset of the BC.

\begin{figure}
 \epsscale{0.75}
 \plotone{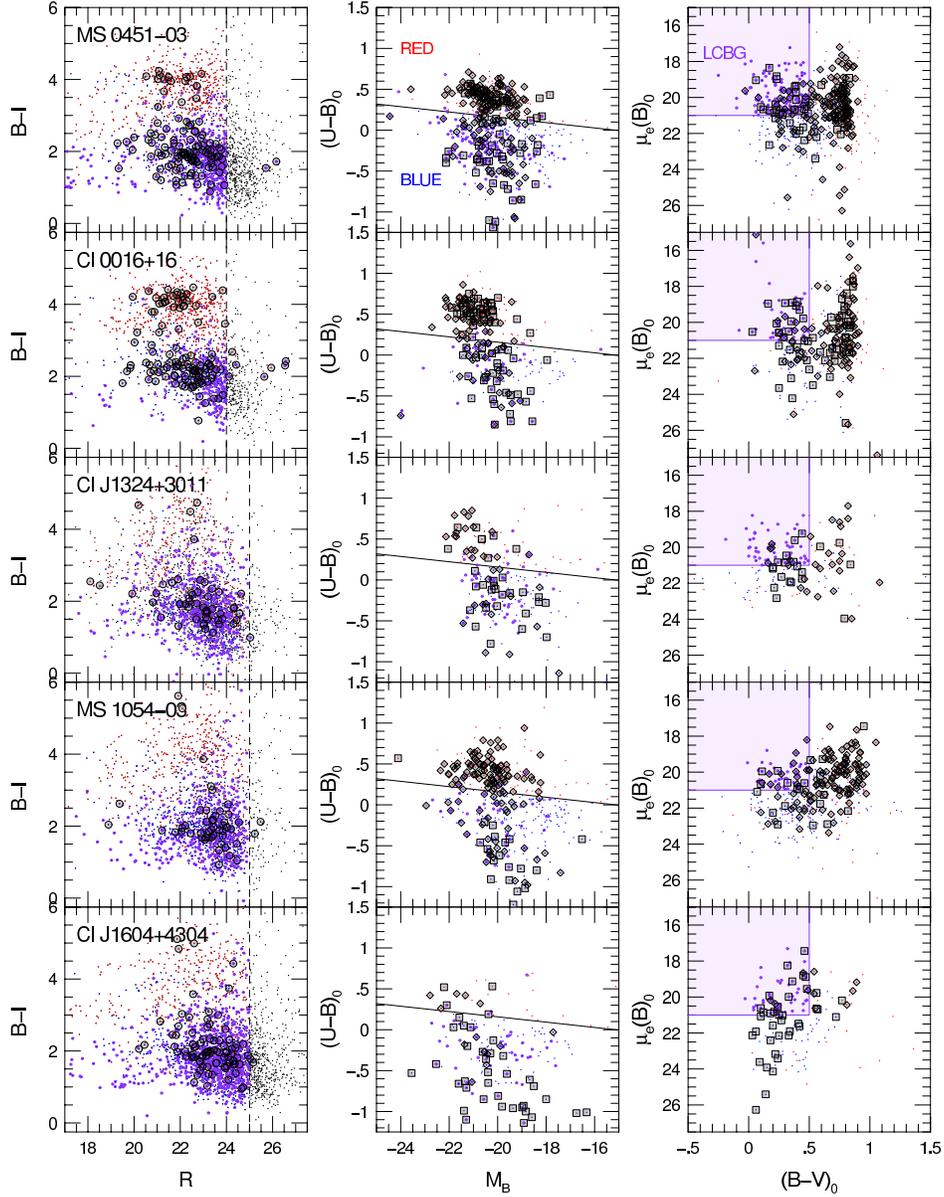}
 \caption{
Photometric properties and sample definitions for objects in our 5
cluster fields.  Each row is for the labeled cluster field. The
left-hand column contains the apparent $B-I$, vs $R$ color-magnitude
diagram. The characteristic sample limit for each cluster is marked by
a vertical dashed line. Grey encircled points have secure
spectroscopic redshifts from our survey. Center and right panels
contain rest-frame $U-B$ and $B-V$ colors, respectively, versus $B$-band
absolute magnitude and rest-frame $B$-band surface brightness within the
half-light radius. Rest-frame quantities are based on best available
redshifts. These panels illustrate the Red Sequence, Blue Cloud, and
LCBG samples, defined in the text, for all galaxies with spectroscopic
redshifts. Grey ensquared points are cluster objects sampled from our
survey; grey diamonds are cluster objects from the literature.
 }
  \label{fig-master-class}
\end{figure}

For the RS class, we adopt the definition originated by Willmer et al.
(2006) and subsequently employed by Crawford et al. (2009).  RS
galaxies are defined as satisfying the following relation:
\begin{equation}
  U-B > -0.032 \times(M_B+21.52)+0.204.
\end{equation}
This definition is based on a $-0.25$ mag shift in the zeropoint of
the color-magnitude relationship at intermediate redshifts.  BC
galaxies are defined as all galaxies below this relationship.  The
distinction between the RS and BC classes is evident in the center
panels of Figure \ref{fig-master-class}.

Finally, the LCBG subset consists of the most compact and luminous
members of the BC class such that they have the following rest-frame
parameters: $(B-V)_o<0.5$, $\mu_B < 21$ \mbox{mag arcsec}$^{-2}$, and
$M_B < -18.5$ (Crawford et al. 2006).  These parameters were defined
to isolate ``enthusiastic'' star forming galaxies, i.e., luminous
galaxies with active star formation ongoing for at least several
hundred million years.  This definition is slightly different than the
one used in Werk \etal\ 2004 and Garland \etal\ 2004, where
$(B-V)_o<0.6$.  The differences between the definitions is minor, and
adopting their definition would only increase our number densities by
$8\%$.  Heavily obscured objects will not be identified as LCBGs.
Figure \ref{fig-master-class} shows our selection for LCBGs in terms
of color and surface brightness.

\section{Redshift Analysis}\label{sxn-analysis}

\subsection{Completeness}\label{sxn-complete}

In total, we attempted to measure spectra from 1288 slits over 15
masks.  Table \ref{tab-zqual} summarizes our overall results. We were
able to measure secure redshifts ($Q=-1$, 3, or 4) for 848 sources,
thus yielding a total success rate of $66\%$.  This is significantly
below the $74\%$ success rate achieved in the Team Keck Redshift
Survey (TKRS, Wirth \etal\ 2004), which used the same instrument with
an exposure time of 3600~s per mask but with a lower-resolution
grating yielding higher signal-to-noise.  In our longer-exposure
masks, we reach a comparable completeness level; thus, our lower
completeness is primarily the result of higher resolution.

In Table \ref{tab-slits}, we list the success rate for determining a
secure redshift ($Q\geq3$) for each of the different masks.  Our
average success rate for the October 2005 run was $69\%$ vs.  $60\%$
for the April 2006 observing run.  The difference between the two runs
can most likely be attributed to the higher redshift of the clusters
in the latter run.  The highest completeness fractions for the second
run occur for the lowest redshift cluster in the group and for the
mask with the longest exposure time.

\subsection{Literature Data}\label{sxn-lit}

\begin{figure}
\plotone{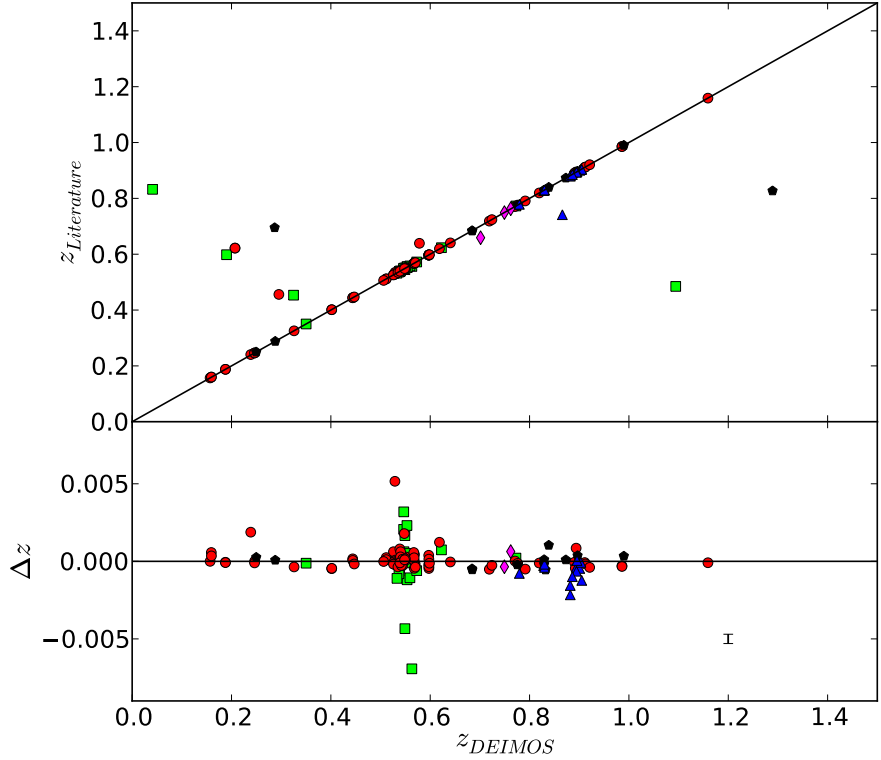}
\caption{\label{fig:zlit}
  Spectroscopic redshifts measured from DEIMOS compared to literature
  sources.  Different clusters are represented by 
red circles (MS 0451-03), 
green squares (Cl 0016+16), 
purple diamonds (Cl J1324+3011), 
black pentagons (MS 1054-03), 
and blue triangles (Cl 1604+4304).  Excluding
  catastrophic outliers, the overall difference of the sample is
  $\Delta z = -0.00013$ with a dispersion of $\sigma_z = 0.0011$. 
  Sources with catastrophic errors are still
  included in this figure.  A typical error bar for the sources
  is included in the bottom-right corner of the bottom panel. }
\end{figure}

Of our 848 galaxies with secure spectroscopic redshifts, 142 have
spectroscopic redshifts from other sources in the literature.  The
vast majority of these redshifts (86 sources) are from DEIMOS
spectroscopy in MS 0451-03 field by Moran et al. (2007)\footnote{Our
selection of targets was done completely independently of the
selection from Moran et al. but the observations were made with the
same instrument and telescope.}.  In Figure \ref{fig:zlit}, we compare
the independent measurements for these 142 sources; only a small
number of significant outliers exist.  We find that $91\%$ have
redshifts that agree to within $|\Delta z| <0.005$, with only 12
measurements that have differences of $|\Delta z | \geq 0.005$.
Excluding the outliers, we find the mean systematic difference between
our redshifts and the literature results to be $\Delta z = -0.00013$
with a dispersion of $\sigma_z = 0.0011$.

\begin{figure}
\plotone{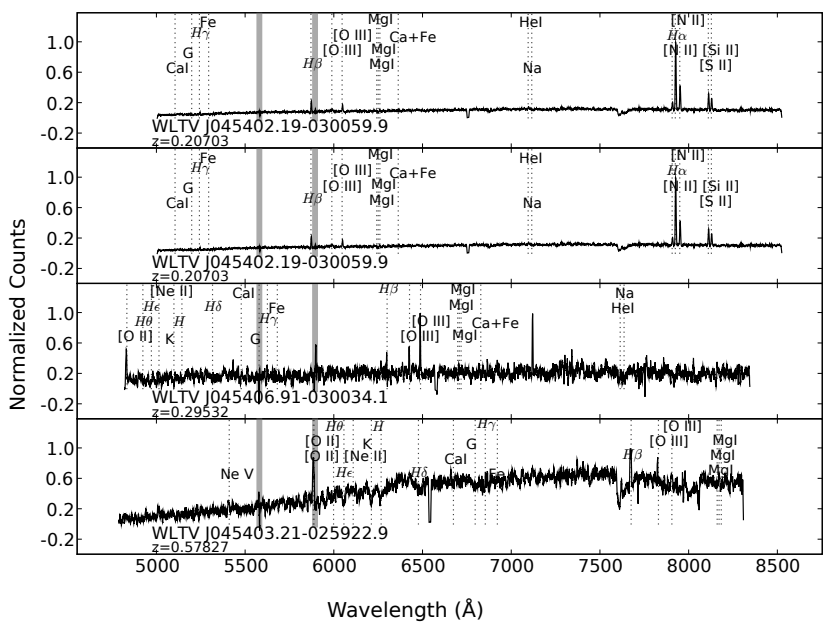}
\caption{\label{fig:spec_w05}
  DEIMOS spectra from the MS 0451-03 field for objects with redshifts
  different from those reported in the literature.  The spectra have
  been smoothed with a boxcar of length 4~\AA\ and important features
  are described in the text.  Grey regions highlight significant sky
  lines and the position of possible detected, individual spectral 
  features are noted in the figure.}
\end{figure}

\begin{figure}
\plotone{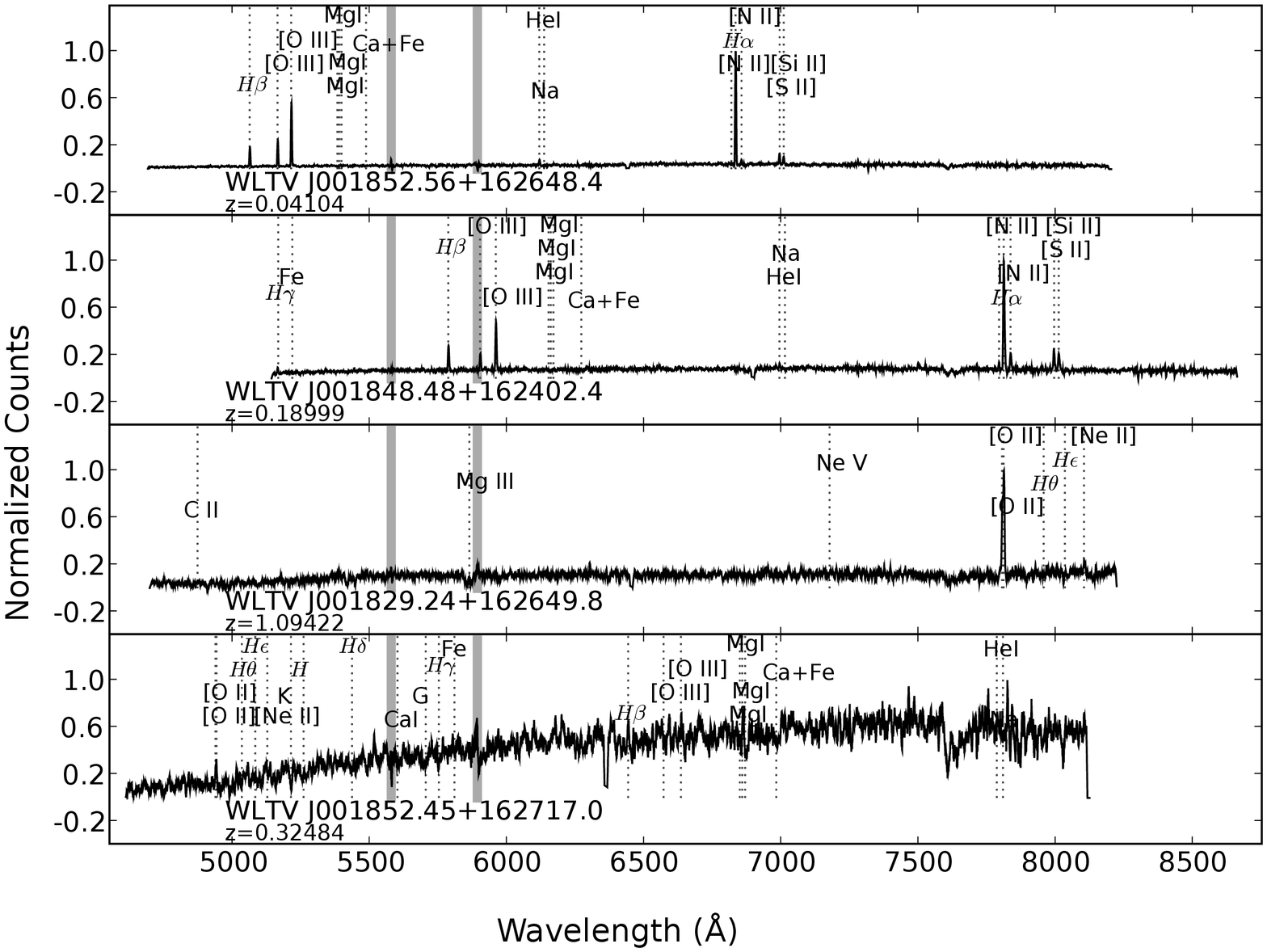}
\caption{\label{fig:spec_w01}
  DEIMOS spectra from the Cl 0016+16 field for objects with redshifts
  different from those reported in the literature.  The spectra have
  been smoothed with a boxcar of length 4~\AA\ and important features
  are described in the text.}
\end{figure}

\begin{figure}
\plotone{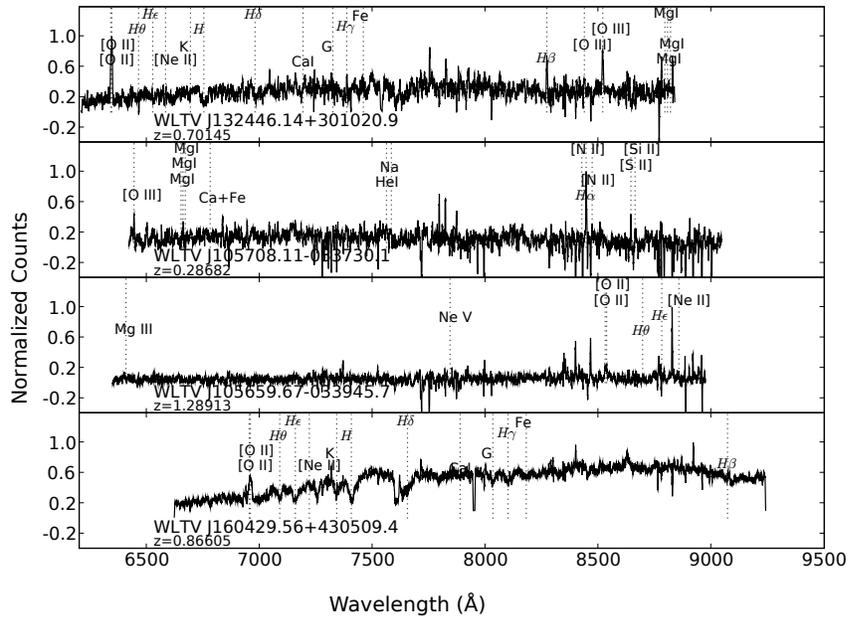}
\caption{\label{fig:spec_oth}
  DEIMOS spectra from the other fields for objects with redshifts
  different from those reported in the literature.  The spectra have
  been smoothed with a boxcar of length 4 ~\AA\ and important features
  are described in the text.}
\end{figure}

The following 11 objects were identified as outliers (one source has
two DEIMOS measurements).  We present all twelve DEIMOS spectra in
Figures \ref{fig:spec_w05}-\ref{fig:spec_oth}.

{\bf WLTV J045402.19-030059.9}: This source was identified as MS
0451.6-0305:PPP 1147 from Ellingson et al. (1998).  The reported
redshift for the source was $z=0.6219$.  We observed this source on
two different DEIMOS masks, and the very secure ($Q=4$) spectroscopic
measurements for this source agree to within $\Delta z = 0.00001$ of
$z=0.20703$.  The presence of H$\beta$, \oiii
$\lambda4959$,$\lambda5007$, and H$\alpha$ confirm the redshift of
this source.  Due to limitations in the wavelength coverage,
resolution, and signal-to-noise of the original spectrum, H$\alpha$
was not identified and \oiii $\lambda5007$ was likely identified as
\oii leading to the erroneous redshift of $z=0.6219$.

{\bf WLTV J045406.91-030034.1}: This source was identified as MS
0451.6-0305:PPP 1349 from Ellingson et al. (1998).  The reported
redshift for the source was $z=0.456$.  We measure a very secure
redshift of $z=0.29532$ based on several emission features.  There is
no obvious reason for the different redshift reported in the
literature source, but the target lies in a crowded region and
confusion is a possibility.

{\bf WLTV J045403.21-025922.9}: This source was identified as MS
0451.6-0305:PPP 1790 by Ellingson et al. (1998).  The reported
redshift for the source was $z=0.6391$ and it was originally
identified as an emission-line source.  We obtained a very secure
redshift measurement of $z=0.57827$ and identify the source as an
absorption line system.  The region is not very crowded, and there is
no evident explanation for the $\Delta z =0.05 $ discrepancy in
redshift.

{\bf WLTV J001852.56+162648.4}: This source was identified as MS
0015.9+1609:PPP 1160 from Ellingson et al. (1998).  Most likely,
H$\alpha$ was misidentified as \oii at $z=0.83226$. We derive a
redshift of $z=0.041038$ due to detecting H$\alpha$ and \oiii
$\lambda5007$ in the spectrum.

{\bf WLTV J001848.48+162402.4}: This source was identified as MS
0015.9+1609:PPP 405 from Ellingson et al. (1998).  Most likely,
H$\alpha$ was misidentified as \oiii $\lambda5007$ at $z= 0.587$.  We
measure the redshift as $z=0.18998$ due to detecting H$\alpha$ and
\oiii $\lambda5007$ in the spectrum.

{\bf WLTV J001829.24+162649.8}: This source was identified as MS
0015.9+1609:PPP 1150 from Ellingson et al. (1998).  The reported
redshift for the source was $z=0.48463$.  We believe the redshift is
$z=1.09422$ based on resolving the \oii doublet which was likely not
visible in the original spectrum.

{\bf WLTV J001852.45+162717.0}: This source was matched with MS
0015.9+1609:PPP 1317 from Ellingson et al. (1998). The reported
redshift for the source was $z=0.4533$.  There was no obvious reason
for the difference between our measured value of $z=0.32484$ and the
value from the literature although we have a very secure ($Q=4$)
measurement of the redshift based on detection of the \oii doublet and
H$\beta$.

{\bf WLTV J132446.14+301020.9}: This source was matched with Cl
J1324+3011 1636 from Postman et al. (2001). The reported redshift was
$z=0.659$.  We detect very strong \oii measured at $z=0.70145.$ There
is no obvious reason for the difference but the source is in a crowded
region and may be misidentified.

{\bf WLTV J105708.11-R033730.1}: This source was matched with MS
J1054-0321 H7758 from Tran et al. (2007).  The reported redshift for
this source was $z=0.6952$.  We measure a secure redshift of
$z=0.28682$ via several emission lines.  It lies in a crowded region
of the field and may be mis-identified.

{\bf WLTV J105659.67-033945.7}: This source corresponds to MS
J1054-0321 K556 from Tran et al. (2007).  There are no strong emission
lines for this object and it is relatively faint at $R=23.45$.  We
measured $z=0.289$ as compared to $z=0.827$ from Tran et al.  In both
surveys, it has a quality of only $Q=3$ and likely a marginal
detection.

{\bf WLTV J160429.56+430509.4}: This source was matched with Cl
J1604+4304 3197 from Postman et al. (2001).  The reported redshift for
this source was $z=0.7415$.  We measure $z=0.86605$.  Our spectrum of
the source reveals a very strong absorption line system, although it
could be blended with another source.

For the eleven discrepant redshifts, seven are from Ellingson et al.
(1998).  As compared to their CFHT MOS observations, the DEIMOS
spectra exhibit higher signal-to-noise, greater wavelength coverage,
and improved resolution.  This allows us to de-blend the \oii doublet
for secure redshift measures as well as to identify other emission
lines out to higher redshift.  Overall, we find that only one of our
redshift measurements among these discrepant sources is marginal,
whereas the other ten are very secure measurements with either the
\oii doublet resolved or multiple lines identified in the spectrum.
Four of the sources are either blended or in crowded regions of the
field and they could be mis-identified with other sources in either
our survey or the previous ones.

\subsection{Accuracy of Photometric Redshifts}\label{sxn-photoz}

We now compare our spectroscopic redshifts to the photometric redshift
measurements from Crawford et al. (2009).  The photometric redshifts
were measured using a hybrid method (Csabai et al. 2003) that combines
the template method (Koo 1985) and training-set method (Connolly et
al. 1995).  After creating a grid of artificial spectral energy
distributions based on the models of Bruzual \& Charlot (2003) that
cover a range of star formation histories, we adjusted the grid in
flux space according to the measured colors of known spectroscopic
sources.  Finally, photometric redshifts were calculated using all of
our flux measurements, including the narrow-band observations, on this
new grid.  Since each cluster was observed with an unique set of
narrow band filters, each cluster has its own unique grid, which was
originally based on the same set of models.  This method corrects for
the effects of incomplete coverage in color space by the models along
with any minor effects introduced by offsets in photometric
calibration.

\begin{figure}
\plotone{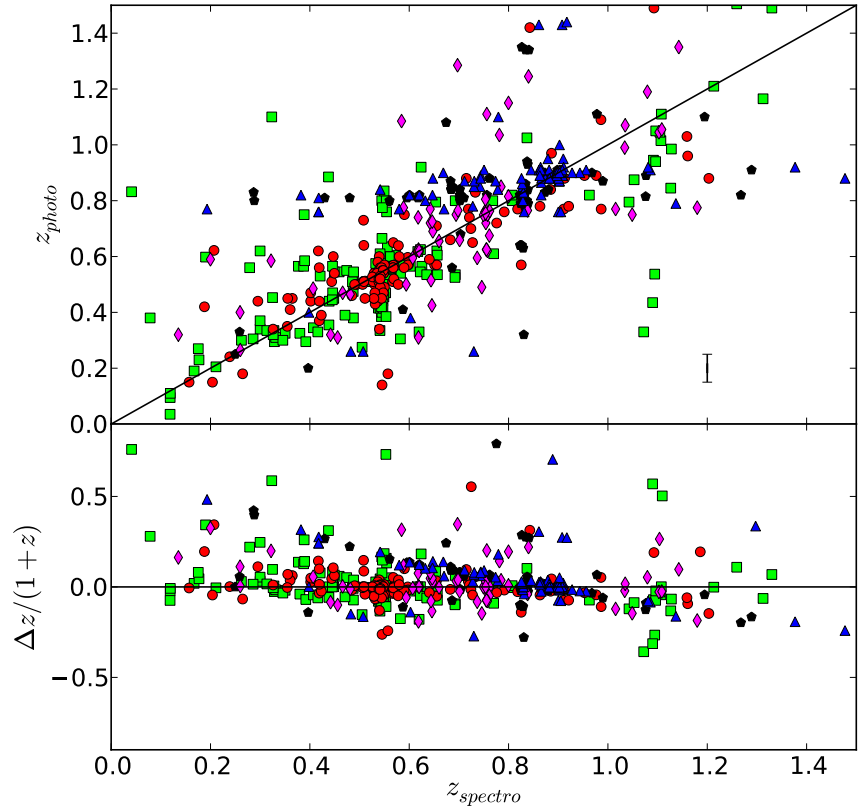}
\caption{\label{fig:zred}
  Photometric redshift compared to spectroscopic redshift excluding
  sources in our original training set data. Symbols are the same as
  in Figure \ref{fig:zlit}.  Sources with catastrophic errors are still
  included in this figure. A typical error bar for the sources
  is included in the bottom-right corner of the top panel. }
\end{figure}

In Figure \ref{fig:zred}, we plot the spectroscopic redshifts versus
the photometric redshifts for all sources with spectroscopic
redshifts, excluding those used in the original training set.  For the
most part, the clusters do not show any major systematic errors and
the overall bias in the sample is relatively small.  The overall
sample has systematic errors of $\delta z/ (1+z) = 0.0015$ and random
error of $\sigma_z /(1+z) = 0.07$ with $10\%$ of the sample being
catastrophic outliers (defined as objects with discrepancies larger
than $3\sigma_z$).  These results are consistent with other studies
(Ilbert et al. 2006, Erben et al. 2009) of similar data quality.  The
sources with the largest errors are those which have been identified
as AGN; we did not include any AGN templates in our original model
grids and thus could not derive accurate redshifts for this class of
galaxies.

One cluster, Cl J1604+4304, does show fairly substantial systematic
errors, especially for lower-redshift sources.  These sources are
predominately faint, blue galaxies that have been assigned photometric
redshifts closer to the cluster redshift than would be appropriate.
After re-examining the training set, we found that this systematic
error resulted from the inclusion of a cluster AGN source in the
training set.  The colors of the AGN were similar to those of
low-redshift blue galaxies and this, along with the small number of
blue galaxies in the original training set, caused the model grid to
be distorted in an unrealistic manner.  However, this highlights a
limitation in this method such that the measured photometric redshifts
are only as good as the training set that is used.  For future
analysis using the photometric redshifts, we plan to recalculate the
model grids using all data now available.

\begin{figure}
\plotone{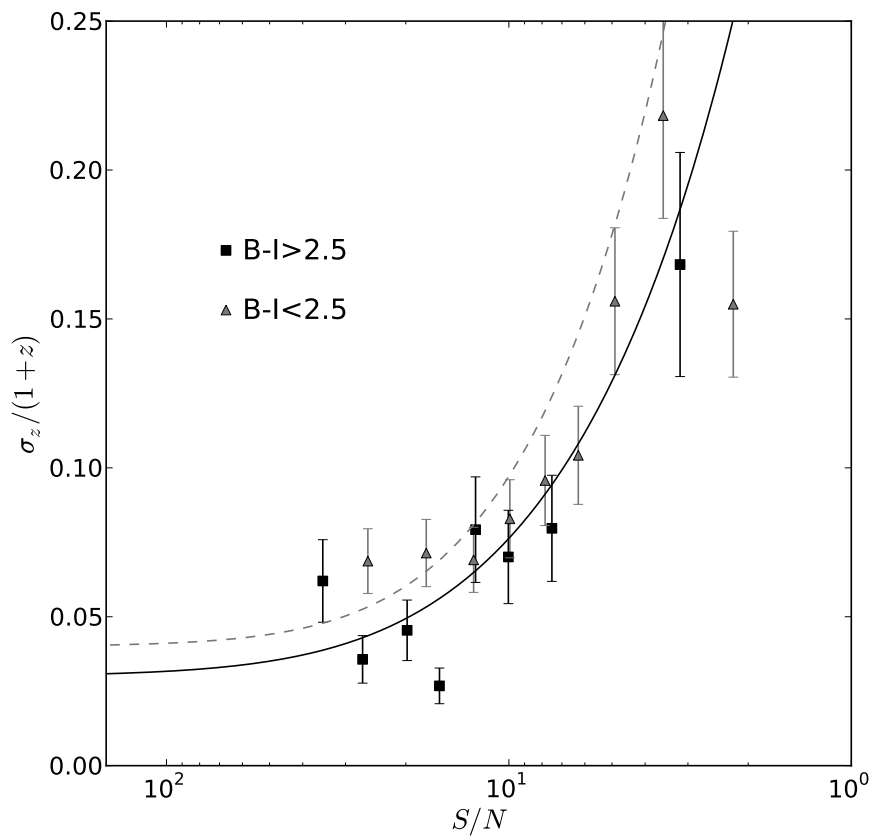}
\caption{\label{fig:zsn}
  Random error in photometric redshifts as a function of signal to
  noise for red ($B-I>2.5$) and blue ($B-I<2.5$) sources.  For red
  sources, each point represents the average of 20 sources; for blue,
  40 sources.  The points are in good agreement with the predictions
  for red (solid line) and blue (dotted line) galaxies based on our
  photometric errors.  }
\end{figure}

To illustrate the importance of photometric errors on the photometric
redshift measurements, we present the random error for the entire
sample as a function of signal-to-noise in the $R$-band in Figure
\ref{fig:zsn}.  We show the data for red and blue objects as defined
by their apparent $B-I$ colors.  For comparison, we plot the expected
random error as a function of signal to noise for two spectral energy
distributions representing a red and blue galaxy assuming photometric
errors typical of our WIYN observations.  Although there is
significant scatter around these models, the data behave as suggested
by the models with a lower limit of $\sigma_z=0.03$ in error for red
sources and $\sigma_z=0.05$ for blue sources and then increasing
rapidly for sources with signal to noise less than 10 in the $R$-band.

\section{Luminous Compact Blue Galaxies in Clusters}
  \label{sxn-lcbgs} 

The initial impetus for this study was to determine the number density
and distribution of LCBGs in intermediate-redshift galaxy clusters.
Crawford \etal\ (2006) found evidence for a large enhancement of the
population of LCBGs using photometric measurements.  Here, we can
confirm their presence with spectroscopic measurements.

\begin{figure}
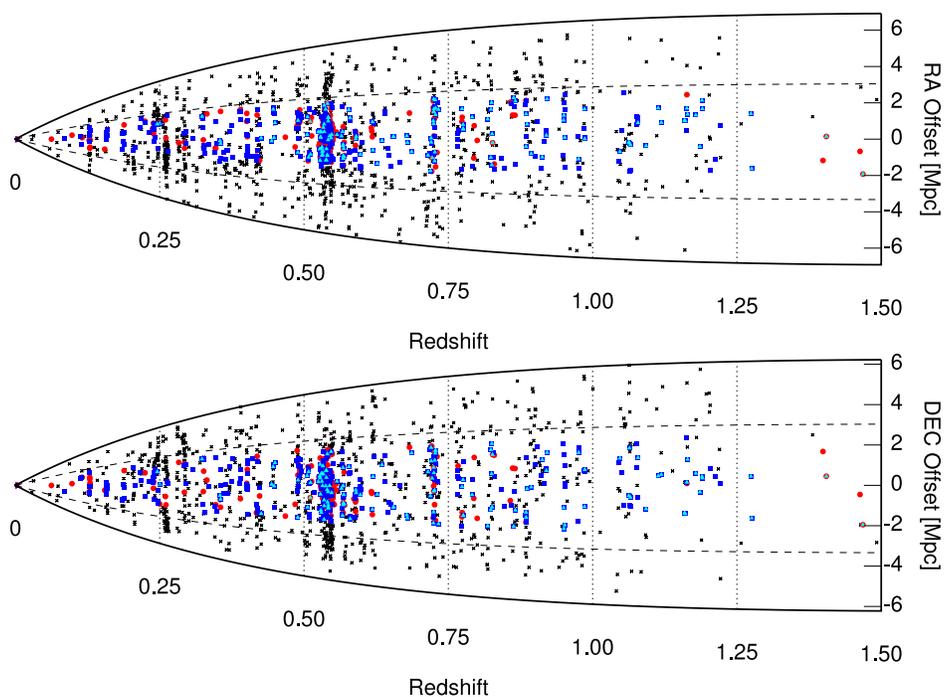

  \plotone{fig13a.epsi}
  \plotone{fig13b.epsi}
  \caption{\label{fig-wedge-w05}
    ``Wedge'' diagrams for the MS0451-03 field.  (a)~Projected spatial
    offset in the RA axis [Mpc] from the cluster center (defined by the
    BCG) among targets in the field.  (b)~Projected spatial offset in
    the Dec axis [Mpc] from the cluster center among targets in the
    field.  Red symbols correspond to galaxies classified as red
    sequence; similarly, blue symbols show the
    blue cloud class, cyan represents the LCBG class, and small
    black crosses indicate additional, unclassified objects with good
    redshifts.  The horizontal dashed curves indicate the extent of
    the DEIMOS survey field.  }
\end{figure}

\begin{figure}
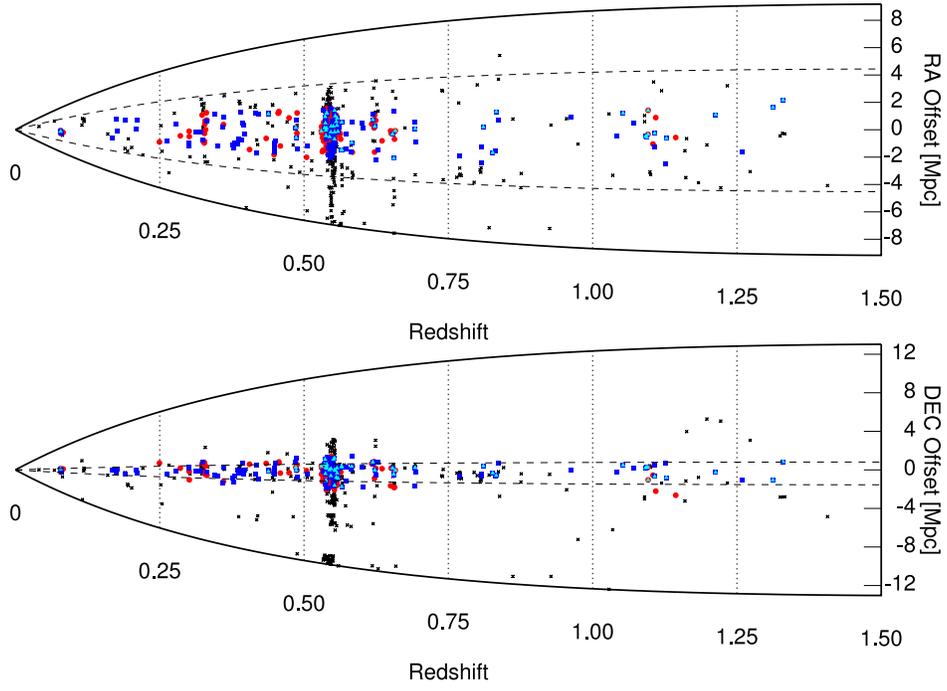

  \plotone{fig14a.epsi}
  \plotone{fig14b.epsi}
  \caption{\label{fig-wedge-w01}
        ``Wedge'' diagrams for the Cl 0016+16 field.  Symbols as in
    Fig.~\ref{fig-wedge-w05}.}
\end{figure}

\begin{figure}
\plotone{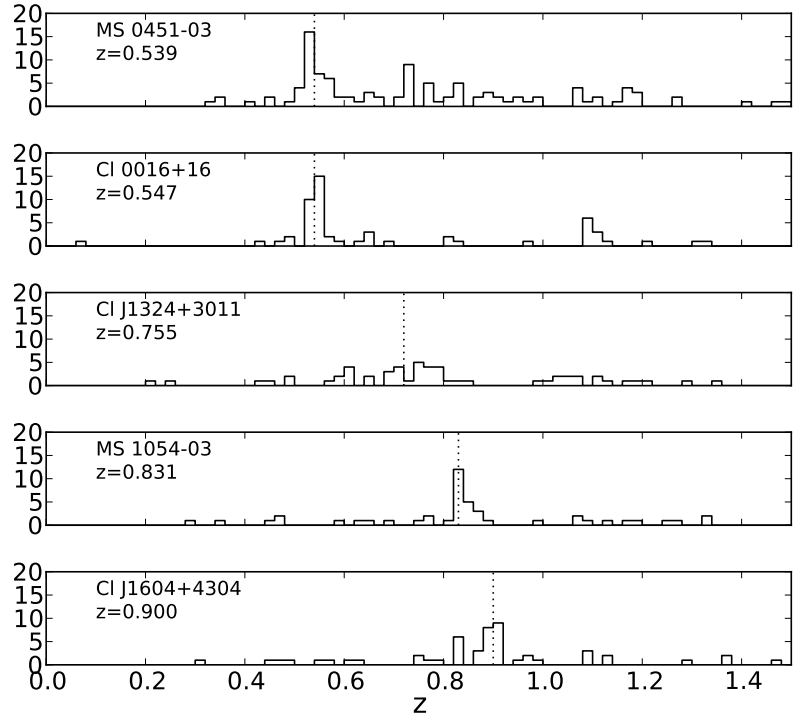}
\caption{\label{fig-zhist} 
 The redshift distribution for LCBGs in the five clusters.  The number
 counts are limited to within $R_{200}$ of each respective cluster
 center.  For all the clusters, an increase in the number of LCBGs is
 noticed  at the cluster redshift (indicated by a dotted vertical line).  }
\end{figure}

In Figures \ref{fig-wedge-w05}-\ref{fig-wedge-w10}, we plot the
spatial distribution of different classes of objects in each of our
five survey fields.  The strong clustering for LCBGs which is implied
in these figures is further demonstrated in the redshift histograms of
LCBGs presented in Figure \ref{fig-zhist}.  A peak in the LCBG
distribution can be seen at the redshift of each cluster with the most
distinct peaks occurring at the more massive clusters.  As shown
previously in Crawford \etal\ (2006), this is further evidence that
the presence of LCBGs correlates with galaxy density.

From our spectroscopy, we identify 145 LCBGs, of which 56 are within
the projected $R_{200}$ radius of the cluster center and $|\Delta z|
\leq 0.03$ of the cluster redshift.  From these measurements, we can
estimate the density of LCBGs within $R_{200}$ of the cluster.  To
account for spectroscopic incompleteness, we estimated the number of
possible cluster LCBGs based on the photometric measurements and
assuming each object was at the redshift of the cluster.  Using the
spectroscopic sources for each cluster, we measured the fraction of
photometrically-determined cluster LCBGs that were bona fide cluster
LCBGs.  For the high redshift cluster, our photometric data includes
all sources within $R_{200}$; for the low redshift clusters, we have
to apply a second correction to account for not sampling the entire
region out to $R_{200}$.  For the two low-redshift clusters, we
calculate the volume within $R_{200}$ being sampled by the photometric
data and correct the density by this fraction. If we assume a volume
for each of our clusters given by a sphere with a radius of $R_{200}$,
we would find a space density for LCBGs in clusters ranging from
$1.65\pm0.25$ Mpc$^{-3}$ at $z=0.55$ to $3.13\pm0.65$ Mpc$^{-3}$ at
$z=0.8$.

In comparison, the field number density of LCBGs also rapidly rises
with redshift with the number density increasing from
$1.2\times10^{-3}$ Mpc$^{-3}$ at $z=0.5$ to $9 \times10^{-3}$
Mpc$^{-3}$ at $z=0.9$ for a similarly defined sample (Phillips \etal\
1997).  Following the same procedure as Phillips \etal\ (1997), we can
calculate the density of field LCBGs over those two redshift ranges by
using our non-cluster sample.  We calculate field densities of
$1.8\pm0.3 \times10^{-3} $Mpc$^{-3}$ at $z=0.5$ and $7.5\pm1.2
\times10^{-3}$ Mpc$^{-3}$ at $z=0.9$ for field LCBGs, which are very
similar to the Phillips \etal\ measurements.  For the low redshift
calculations, we used the Cl J1324+3011, MS 1054-03, and Cl J1604+4304
fields; for the high redshift, MS 0451-03 and Cl 0016+16 fields.  We
adopt the Phillips \etal\ (1997) values due to the better
spectroscopic completeness in their data for the high redshift field
samples; however, this choice does not significantly change our
results presented here.

\begin{figure}
\plotone{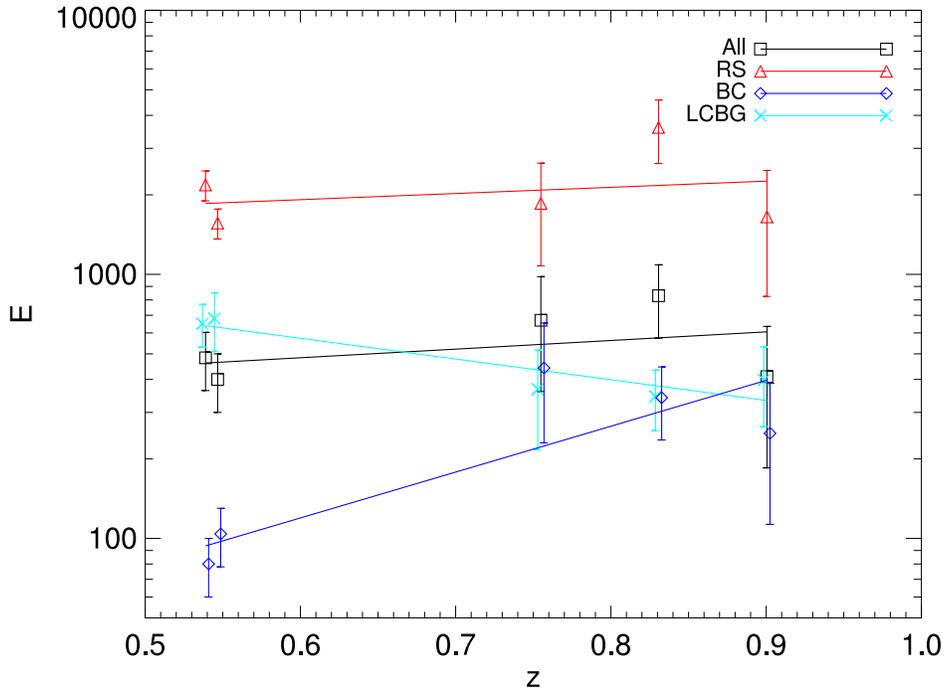}
\caption{\label{fig-enhance}
Enhancement factor $E$ as a function of redshift $z$ for the five
clusters in our sample, for red sequence galaxies (RS), blue cloud
objects (BC), luminous compact blue galaxies (LCBG), and all types
together.  The value of $E$ is defined as the relative density of a
given galaxy type in the cluster vs. the field at a given redshift.
The indicated line for each class represents a linear fit to $\log(E)$
as a function of $z$, accounting for the measurement errors.  The RS
class shows little evolution in $E$ within our sample.  The relative
BC population density rises significantly with increasing redshift.
The corresponding LCBG population density \emph{relative to the field}
displays the opposite behavior, decreasing strongly as redshift
increases and thus suggesting strong \emph{differential} evolution in
the BC and LCBG populations.}
\end{figure}

Since the cluster space density depends on the richness of the
selected cluster, to make sense of the differences between the cluster
and the field we compute an enhancement, $E$, defined as the ratio of
cluster to field density.  LCBGs have an average enhancement of $E=749\pm116$
at $z=0.55$ and $E=416\pm95$ at $z=0.8$.  For comparison, the
enhancement of red sequence galaxies can be calculated from their
field (Willmer \etal\ 2006) and cluster (Crawford et al. 2009)
luminosity functions.  At $z=0.55$ ($0.80$), red sequence galaxies have an
enhancement of $E=1872\pm174$ ($2369\pm500$).  Using the measured blue
fraction in each cluster, we can estimate the enhancement of BC
galaxies to be $E=92\pm12$ ($344\pm78$) and for all types of galaxies
to be $E=440\pm78$ ($636\pm173$) at $z=0.55$ ($0.8$), respectively.  The
enhancement for RS, BC, and LCBGs for each cluster are given in Table
\ref{tab_enhance}, and the change with redshift of the enhancement for 
each class of objects can be seen in Figure \ref{fig-enhance}.

At intermediate redshifts of $z=0.5$, these results indicate that
LCBGs are preferentially found in high-density environments relative
to the overall star-forming population.  Although they are not as
strongly clustered as red galaxies (by a factor of 3), they are $1.5$
times as clustered as the overall galaxy distribution and seven times
more clustered than regular blue cloud galaxies in general.  At higher
redshifts, LCBGs also had a high density of objects in clusters, but
the field density of LCBGs was significantly higher (Guzman \etal\
1997).  This results in a factor of two lower LCBG enhancement at this
earlier epoch.  This is remarkable because the enhancement of blue
galaxies is a factor of 3.5 higher at $z=0.8$, thereby equalizing the
enhancement of blue and LCBG populations at a redshift where there is
a large fraction of LCBGs in the blue field population.  In other
words, there is a very strong differential evolution of subsets of the
blue galaxy population between clusters and the field between a
redshift of $z=0.8$ to $z=0.5$.

Cluster blue galaxies are assumed to be an infalling field population
which is extinguished by different processes in the cluster (Dressler
\etal\ 1997; Balogh, Navarro, \& Morris 2000; Ellingson \etal\ 2001,
Bravo-Alfaro \etal\ 2001, Chung \etal\ 2009).  If we assume the same
is true for LCBGs, we would expect their number density to follow a
similar pattern to either the overall population or the BC population,
which is the case for our higher redshift clusters.  For the
$z\sim0.55$ clusters, we find a much higher number of LCBGs than we
would predict from this simple model. In these clusters, LCBGs are
completely absent in the very high-density cores of the clusters
(Crawford \etal\ 2006).  This is strong evidence that the cluster
environment is triggering the starburst in these galaxies (Porter
\etal\ 2008; Mahajan, Haines, and Raychaudhury 2011).

In the original work (Crawford \etal\ 2006), we couched the
enhancement as a way to connect different galaxy populations by their
morphology-density relationship.  Low-redshift dwarf spheroidal
galaxies, which Koo \etal\ (1994) originally proposed as a possible
descendant of LCBGs, have a similar enhancement to the
intermediate-redshift LCBGs.  However, the higher-redshift LCBGs are
likely a more heterogeneous population and their evolutionary path is
more complex (Phillips \etal\ 1997, Noeske et al. 2006).  Hence, more
detailed information about the individual objects will be needed to
connect these cluster objects with their lower-redshift cluster
relatives.  Nonetheless, our findings here on enhancement bear out our
earlier results based on photometric redshifts alone.

Unlike field LCBGs (Guzman \etal\ 1997), cluster LCBGs show only a
modest decrease in number density with redshift.  As compared to RS
and BC galaxies, LCBGs are the only group that shows an increase in
the enhancement with decreasing redshift as seen in Figure
\ref{fig-enhance}. This study shows that massive clusters at
intermediate redshifts still contain a relative abundance of LCBGs
despite their increasing rarity in the field, perhaps because the
cluster periphery is a fertile environment for triggering the LCBG
phase in in-falling gas-rich galaxies.  The clusters in our survey are
representative of the most massive systems in the Universe.  As seen
in the comparison between Abell 851 and MS 1512.4+3647 (Lotz \etal\
2003), the extreme environment in massive systems may lead to very
different properties than the more common, lower mass systems.  It
will be important to broaden this type of investigation to a range of
environments and redshifts to further explore the triggering of LCBGs.
Furthermore, confirmation of this trend is still required at low
redshifts where field LCBGs are almost non-existent (Werk \etal\
2004).  Future studies targeting the periphery of low-redshift, rich
galaxy clusters could confirm whether this trend continues to today.

\section{Summary}\label{sxn-summary}

We have presented the spectroscopic observations of blue galaxies in
five moderate-to-high redshift galaxy clusters.  The five clusters
targeted here include some of the most massive systems at their
respective redshifts and we have preferential targeted blue sources
associated with the clusters.  This paper is the first in a series
attempting to determine a complete census of the properties of optical
star-forming galaxies in intermediate-redshift galaxy clusters.

We have detailed the DEIMOS spectroscopic observations for blue
galaxies selected from a deep, multi-band imaging survey with the WIYN
3.5~m telescope.  This includes the object selection, observations,
data reduction, and analysis.  In addition, we present a table of the
measurements for all 1288 sources that were targeted as part of this
survey including spectroscopic redshift and photometric
classification.

We determined secure redshifts for 848 sources. Our success rate for
determining the redshift for sources is comparable to previous studies
with the same instrument and telescope.  In our sample, 142 sources
have redshifts previously reported in the literature.  Twelve
measurements (11 sources) are discrepant with the literature values,
although redshifts are very securely ($Q=4$) determined for ten of
these sources.  Overall, our results show excellent agreement with the
previously-published results.  Comparing the spectroscopic redshifts
to our previously-measured photometric measurements yields results
that confirm the high quality of our photometric measurements.
Photometric redshifts from one cluster did exhibit systematic errors
for low-redshift blue sources, which we attribute to AGN contamination
in the original training set.  The overall dispersion in the
measurement is comparable to our expectations from modeling our
photometric errors.

We have estimated the number density of LCBGs in these five galaxy
clusters based on our new spectroscopic identifications.  By examining
the number distribution of LCBGs as a function of redshift, we find
the clusters to be rich in LCBGs with a relative enhancement over the
field population of a factor of 500, roughly 2.5 times larger than the
enhancement of the general blue cluster population.  The relative
enhancement between sub-populations of star-forming galaxies diverges
between $z=0.8$ to $z=0.5$ such that LCBGs become relatively more
common in massive clusters at more recent epochs.  This overdensity of
luminous compact star-forming galaxies indicates that the cluster
environment, while generally accelerating the transformation of
galaxies from the blue cloud to the red sequence, is somehow better
able to nurture or sustain the LCBG phase relative to the field.




\acknowledgments


We thank the referee for the careful reading of our manuscript and the
constructive criticism that improved our paper.  We wish to thank John
Hoessel for his early contributions to the imaging survey and
selection of the the cluster fields, Keck Observing Assistants Carolyn
Parker and Joel Aycock for helping with the Keck observations, and
Daniel Gregg for assisting with the redshift measurements.  SMC and
MAB wish to acknowledge STScI/AR-9917. SMC wishes to acknowledge the
South African Astronomical Observatory and the National Research
Foundation of South Africa for support during this project.  MAB
acknowldges support from NSF grant AST-1009491.

The authors wish to recognize and acknowledge the very significant
cultural role and reverence that the summit of Mauna Kea has always
had within the indigenous Hawaiian community.  We are most fortunate to
have the opportunity to conduct observations from this mountain.  The
analysis pipeline used to reduce the DEIMOS data was developed at UC
Berkeley with support from NSF grant AST-0071048. This research has
made use of the NASA/IPAC Extragalactic Database (NED) which is
operated by the Jet Propulsion Laboratory, California Institute of
Technology, under contract with the National Aeronautics and Space
Administration.


{\it Facilities:} \facility{WMKO} \facility{WIYN}

\clearpage

\begin{landscape}
\begin{deluxetable}{llllrrrrr}
  \tablewidth{0pc}
  \tablecaption{Summary of Fields\label{tab_cluster_params}}
  \tablehead{
    \colhead{Field}         &
    \colhead{WLTV ID\tablenotemark{a}}         &
    \colhead{$\alpha$\tablenotemark{b}} &
    \colhead{$\delta$\tablenotemark{b}} &
    \colhead{$z$\tablenotemark{c}} &
    \colhead{$\sigma(z)$\tablenotemark{d}} &
    \colhead{$M_{200}$\tablenotemark{e}} &
    \colhead{$\Rvirial$\tablenotemark{f}} &
    \colhead{$\Rvirial$\tablenotemark{g}} \\
    &
    &\colhead{(J2000)}
    &\colhead{(J2000)}
    &
    &\colhead{(\kms)}
    &\colhead{($10^15 M_{\odot}$)}
    &\colhead{(Mpc)}
    &\colhead{($\arcsec$)}
  }
  \startdata
MS 0451-03 & w05 & 04:54:10.8 & $-$03:00:51 & 0.5389 & 1328 & 3.00 & 2.45 & 386 \\
Cl 0016+16 & w01 & 00:18:33.6 & $+$16:26:16 & 0.5467 & 1490 & 4.22 & 2.74 & 428 \\ 
Cl J1324+3011 & w08 & 13:24:48.8 & $+$30:11:39 & 0.7549 & 806 & 0.59 & 1.31 & 178 \\
MS 1054-03 & w07 & 10:56:60.0 & $-$03:37:36 & 0.8307 & 1105 & 1.45 & 1.72 & 225 \\
Cl J1604+4304 & w10 & 16:04:24.0 & $+$43:04:39 & 0.9005 & 1106 & 1.40 & 1.65 & 211 \\
  \enddata
  \tablenotetext{a}{Internal designation for each of the clusters.}
  \tablenotetext{b}{Celestial coordinates of the adopted cluster center defined by Brightest Cluster Galaxy.}
  \tablenotetext{c}{Measured cluster redshift.}
  \tablenotetext{d}{Measured cluster velocity dispersion.}  
  \tablenotetext{e}{Cluster virial mass computed from $\sigma(z)$.}  
  \tablenotetext{f}{Cluster virial radius computed from $\sigma(z)$.} 
  \tablenotetext{g}{Cluster virial radius in angular units for our
    adopted cosmology.} 
\end{deluxetable}
\end{landscape}


\begin{deluxetable}{rcrrr}
  \tablewidth{0pc}
  \tablecaption{Slitmask Design Data\label{tab-maskdesign}}
  \tablehead{
    \colhead{No.}         &
    \colhead{Mask Name}         &
    \colhead{$\alpha$\tablenotemark{a}} &
    \colhead{$\delta$\tablenotemark{a}} &
    \colhead{PA\tablenotemark{b}} \\
    &&\colhead{(J2000)}&\colhead{(J2000)}&\colhead{($\degr$)}}
  \startdata
 1 & w01.m1 & 00~18~33.63 &    16~26~30.0 & 270  \\
 2 & w01.m2 & 00~18~33.63 &    16~26~30.0 & 270  \\
 3 & w01.m3 & 00~18~33.63 &    16~26~30.0 & 270  \\
 4 & w01.m4 & 00~18~33.63 &    16~26~30.0 & 270  \\
 5 & w05.m1 & 04~54~10.81 & $-$03~00~56.9 & 45   \\
 6 & w05.m2 & 04~54~10.81 & $-$03~00~56.9 & 45   \\
 7 & w05.m3 & 04~54~10.81 & $-$03~00~56.9 & 315  \\
 8 & w05.m4 & 04~54~10.81 & $-$03~00~56.9 & 315  \\
 9 & w07.m1 & 10~56~59.09 & $-$03~38~02.8 & 41  \\
10 & w07.m3 & 10~57~03.33 & $-$03~36~54.2 & 320  \\
11 & w08.m1 & 13~25~03.53 &    30~10~54.0 & 270  \\
12 & w08.m2 & 13~25~03.53 &    30~10~54.0 & 270  \\
13 & w10.m1 & 16~04~21.16 &    43~04~10.9 & 41  \\
14 & w10.m2 & 16~04~19.26 &    43~04~05.2 & 41  \\
15 & w10.m3 & 16~04~20.40 &    43~04~57.0 & 320  \\
  \enddata
\tablenotetext{a}{Celestial coordinates of the nominal slitmask center.}
\tablenotetext{b}{Position angle of the slitmask. }
\end{deluxetable}



\begin{deluxetable}{rrcrrrrrc}
  \tablewidth{0pc}
  \tablecaption{Slitmask Observation Data\label{tab-obslog}}
  \tablehead{
    \colhead{No.}         &
    \colhead{Mask}         &
    \colhead{Obs. Date}         &
    \colhead{Int. Time}         &
    \colhead{Grating}         &
    \colhead{Blaze}         &
    \colhead{Filter}         &
    \colhead{$\lambda_c$}         &
    \colhead{$\lambda$ range\tablenotemark{a}} \\
    &&\colhead{(UT)}&\colhead{(s)}&\colhead{(l~mm\inv)}&\colhead{(\AA)}&&\colhead{(\AA)}&\colhead{(\AA\AA)}}
  \startdata
 1 & w01.m1 & 2005 Nov 03 & 4800 & 900 & 5500 & GG455 & 6500 & 4700--8300  \\
 2 & w01.m2 & 2005 Nov 03 & 3600 & 900 & 5500 & GG455 & 6500 & 4700--8300  \\
 3 & w01.m3 & 2005 Nov 03 & 3600 & 900 & 5500 & GG455 & 6500 & 4700--8300  \\
 4 & w01.m4 & 2005 Nov 03 & 3600 & 900 & 5500 & GG455 & 6500 & 4700--8300  \\
 5 & w05.m1 & 2005 Nov 03 & 3600 & 900 & 5500 & GG455 & 6500 & 4700--8300  \\
 6 & w05.m2 & 2005 Nov 03 & 3600 & 900 & 5500 & GG455 & 6500 & 4700--8300  \\
 7 & w05.m3 & 2005 Nov 03 & 3600 & 900 & 5500 & GG455 & 6500 & 4700--8300  \\
 8 & w05.m4 & 2005 Nov 03 & 3600 & 900 & 5500 & GG455 & 6500 & 4700--8300  \\
 9 & w07.m1 & 2007 Apr 17 & 3600 & 1200 & 7760 & OG550 & 7800 & 6450--9150 \\
10 & w07.m3 & 2007 Apr 17 & 3600 & 1200 & 7760 & OG550 & 7800 & 6450--9150 \\
11 & w08.m1 & 2007 Apr 17 & 3600 & 1200 & 7760 & OG550 & 7500 & 6150--8850 \\
12 & w08.m2 & 2007 Apr 17 & 3600 & 1200 & 7760 & OG550 & 7500 & 6150--8850 \\
13 & w10.m1 & 2007 Apr 17 & 3600 & 1200 & 7760 & OG550 & 8000 & 6650--9350 \\
14 & w10.m2 & 2007 Apr 17 & 3600 & 1200 & 7760 & OG550 & 8000 & 6650--9350 \\
15 & w10.m3 & 2007 Apr 17 & 4800 & 1200 & 7760 & OG550 & 8000 & 6650--9350 \\
  \enddata
\tablenotetext{a}{Nominal wavelength range for a slit lying in the center of the mask; actual wavelength range depends on slit position.}
\end{deluxetable}

\begin{deluxetable}{rrrrrrrrrrr}
\tabletypesize{\tiny}\tablewidth{0pc}
\tablecaption{WLTV DEIMOS Catalog\tablenotemark{a}
\label{tab-data}}
\tablecaption{WLTV DEIMOS Catalog\tablenotemark{a} \label{tab-data}}
\tablehead{
\colhead{ID} & 
\colhead{$\alpha$} & 
\colhead{$\delta$} & 
\colhead{$R$} & 
\colhead{Mask} & 
\colhead{Slit} & 
\colhead{$z$} & 
\colhead{$Q$} & 
\colhead{$z_{\mbox{lit}}$} & 
\colhead{Ref\tablenotemark{b}} & 
\colhead{Class} \\ 
\colhead{} & 
\colhead{(J2000)} & 
\colhead{(J2000)} & 
\colhead{(mag)} & 
\colhead{} & 
\colhead{} & 
\colhead{} & 
\colhead{} & 
\colhead{} & 
\colhead{} & 
\colhead{} \\
\colhead{(1)} & 
\colhead{(2)} &
\colhead{(3)} &
\colhead{(4)} &
\colhead{(5)} &
\colhead{(6)} &
\colhead{(7)} &
\colhead{(8)} &
\colhead{(9)} &
\colhead{(10)} &
\colhead{(11)} 
}
\startdata 
 WLTV J045352.64-030353.4 &   73.4693271  &   -3.0648573 & 19.28 & w05.m1 &   0 &  0.41676 &  4 &      ... & ... &   BC \\ 
 WLTV J045355.35-030636.1 &   73.4806362  &   -3.1100513 & 19.54 & w05.m1 &   2 &  0.25792 &  4 &      ... & ... &   RS \\ 
 WLTV J045353.13-030625.5 &   73.4713687  &   -3.1071082 & 22.32 & w05.m1 &   3 &  0.89036 &  4 &  0.89030 &   5 &   BC \\ 
 WLTV J045353.84-030600.8 &   73.4743266  &   -3.1002444 & 21.76 & w05.m1 &   4 &  0.54157 &  4 &      ... & ... &   BC \\ 
 WLTV J045354.32-030451.0 &   73.4763465  &   -3.0808360 & 21.27 & w05.m1 &   5 &  0.58863 &  3 &      ... & ... &   BC \\ 
 WLTV J045354.48-030518.1 &   73.4769991  &   -3.0883711 & 23.91 & w05.m1 &   6 &  0.77239 &  2 &      ... & ... &   BC \\ 
 WLTV J045355.23-030555.3 &   73.4801383  &   -3.0986969 & 21.98 & w05.m1 &   7 &  0.56637 &  4 &      ... & ... &   BC \\ 
 WLTV J045356.11-030346.2 &   73.4837898  &   -3.0628533 & 21.39 & w05.m1 &   8 &  0.56693 &  4 &  0.56750 &   2 &   BC \\ 
\enddata 
\
\tablecomments{(1) Identification in WLTV survey (2) Right Ascension (3) Declination (4) R mag (5) Mask (6) Slit (7) Redshift (8) Redshift quality (9) Literature redshift (10) Reference (11) Photometric classification. } 
\tablenotetext{a}{A full version of this catalog appears in the electronic edition.}
\tablenotetext{b}{\ List of References: (1) Dressler \& Gunn 1989 (2) Ellingson et al. 1998 (3) Postman et al. 2001 (4) Nakamura et al. 2006 (5) Moran et al. 2007 (6) Tanaka et al. 2007 (7) Tran et al. 2007} 
\end{deluxetable}


\begin{deluxetable}{rp{0.45\textwidth}rr}
  \tablewidth{0pc}
  \tablecaption{Redshift Quality Distribution\label{tab-zqual}}
  \tablehead{
    \colhead{$Q$\tablenotemark{a}}                         &
    \colhead{Definition}                         &
    \colhead{$N_{obj}$\tablenotemark{b}}                   &
    \colhead{$F_{obj}$\tablenotemark{c}}              
  }
  \startdata
  4 &  Very secure redshift ($P>99\%$); at least two
  spectral features identified & 550 & 0.428 \\
  3 &  Secure redshift ($P>95\%$); one strong line and
  another weak feature identified \emph{or} single wide line
  & 177 & 0.138 \\ 
  2 &  Uncertain redshift; signal is present but no unambiguous
  spectral line identified & 85 & 0.066 \\
  1 & No redshift; $S/N$ too poor & 290 & 0.225 \\
  $-1$ & Star & 121 & 0.094 \\
  $-2$ & No redshift measured because of instrumental artifacts in
  spectrum  & 56 & 0.043 \\
  \enddata
\tablenotetext{a}{Redshift quality category.}
\tablenotetext{b}{Number of objects in catalog for this category.}
\tablenotetext{c}{Fraction of targetted objects for this category.}
\end{deluxetable}



\begin{deluxetable}{rrrrr}
  \tablewidth{0pc}
  \tablecaption{Slitmask Results\label{tab-slits}}
  \tablehead{
    \colhead{No.}         &
    \colhead{Mask Name}         &
    \colhead{$N_o$\tablenotemark{a}}    &
    \colhead{$N_z$\tablenotemark{b}} &
    \colhead{$F_z$\tablenotemark{c}} \\
    &&&&\colhead{(\%)}}
  \startdata
 1 & w01.m1  & 93 &  69 & 74 \\
 2 & w01.m2  & 90 &  59 & 67 \\
 3 & w01.m3  & 89 &  58 & 65 \\
 4 & w01.m4  & 87 &  57 & 66 \\
 5 & w05.m1  & 93 &  67 & 73 \\
 6 & w05.m2  & 90 &  52 & 58 \\
 7 & w05.m3  & 90 &  70 & 77 \\
 8 & w05.m4  & 89 &  64 & 71 \\
 9 & w07.m1  & 79 &  40 & 51 \\
10 & w07.m3  & 64 &  31 & 48 \\
11 & w08.m1  & 87 &  58 & 68 \\
12 & w08.m2  & 85 &  65 & 77 \\
13 & w10.m1  & 82 &  49 & 61 \\
14 & w10.m2  & 72 &  35 & 50 \\
15 & w10.m3  & 73 &  48 & 66 \\
  \enddata
\tablenotetext{a}{Number of objects per mask; note that a slit may
  contain multiple objects.}
\tablenotetext{b}{Number of secure redshifts ($Q=-1$, 3, or 4) measured per mask.}
\tablenotetext{c}{Percentage of objects per mask yielding secure redshifts.}
\end{deluxetable}



\begin{deluxetable}{lrrrrr}
  \tablewidth{0pc}
  \tablecaption{Enhancement in each cluster\tablenotemark{a}\label{tab_enhance}}
  \tablehead{
    \colhead{Cluster}         &
    \colhead{z}         &
    \colhead{All}         &
    \colhead{RS}    &
    \colhead{BC} &
    \colhead{LCBG} \\
  }
  \startdata
 MS 0451-03    & $0.5389$ & $483\pm120$ & $2180\pm283$ & $80\pm20$  & $786\pm156$ \\
 Cl 0016+16    & $0.5467$ & $400\pm100$ & $1564\pm203$ & $104\pm26$ & $712\pm174$ \\
 Cl J1324+3011 & $0.7549$ & $670\pm310$ & $1857\pm780$ & $442\pm212$ & $574\pm246$ \\
 MS 1054-03    & $0.8307$ & $830\pm257$ & $3600\pm972$ & $341\pm105$ & $326\pm78 $ \\
 Cl J1604+4304 & $0.9005$ & $410\pm225$ & $1650\pm825$ & $250\pm137$ & $348\pm122$ \\
  \enddata
\tablenotetext{a}{The enhancement is defined as the ratio between the cluster and field density}
\end{deluxetable}

\clearpage

\begin{figure}
  \caption{Projected sky distribution for targets in the MS 0451-03
    field.  Dotted box indicates the approximate field-of-view of the
    WIYN photometry from which the photometric redshifts were derived.
    Solid polygon indicates the field-of-view of the DEIMOS
    spectroscopy from this work.  Dashed circle indicates the
    $R_{200}$ radius for this cluster.  Green pluses represent objects
    with WIYN photometry.  Cyan circles denote targets with
    previously-published redshifts appearing in the literature.  Red
    crosses specify objects targeted with DEIMOS that did not yield a
    secure redshift measurement.  Blue diamonds correspond to
    galaxies with secure DEIMOS redshifts.  Yellow symbols identify stars.
    The figure is only available in high-resolution version.
    \label{fig-radec-dss-w05}
  }
\end{figure}

\begin{figure}
  \caption{Projected sky distribution for targets in Cl 0016+16
    field.  Symbols as in Fig.~\ref{fig-radec-dss-w05}.
    The figure is only available in high-resolution version.
    \label{fig-radec-dss-w01}
  }
\end{figure}

\begin{figure}
  \caption{Projected sky distribution for targets in Cl J1324+3011
    field.  Symbols as in Fig.~\ref{fig-radec-dss-w05}.
    The figure is only available in high-resolution version.
    \label{fig-radec-dss-w08}
  }
\end{figure}

\begin{figure}
  \caption{Projected sky distribution for targets in MS 1054-03
    field.  Symbols as in Fig.~\ref{fig-radec-dss-w05}.
    The figure is only available in high-resolution version.
    \label{fig-radec-dss-w07}
  }
\end{figure}

\begin{figure}
  \caption{Projected sky distribution for targets in Cl J1604+4304
    field.  Symbols as in Fig.~\ref{fig-radec-dss-w05}.
    The figure is only available in high-resolution version.
    \label{fig-radec-dss-w10}
  }
\end{figure}

\begin{figure}
  \plotone{fig15a.epsi}
  \plotone{fig15b.epsi}
  \caption{\label{fig-wedge-w08}
        ``Wedge'' diagrams for the w08 field.  Symbols as in
    Fig.~\ref{fig-wedge-w05}.}
\end{figure}

\begin{figure}
  \plotone{fig16a.epsi}
  \plotone{fig16b.epsi}
  \caption{\label{fig-wedge-w07}
        ``Wedge'' diagrams for the w07 field.  Symbols as in
    Fig.~\ref{fig-wedge-w05}.}
\end{figure}

\begin{figure}
  \plotone{fig17a.epsi}
  \plotone{fig17b.epsi}
  \caption{\label{fig-wedge-w10}
        ``Wedge'' diagrams for the w10 field.  Symbols as in
    Fig.~\ref{fig-wedge-w05}.}
\end{figure}

\clearpage

\clearpage

\end{document}